\documentclass[11pt,preprint]{aastex}
\usepackage{times}

\def\etal{et al.\rm}

\lefthead{W. Colley and R. Schild}
\righthead{Q0957 Rapid Microlensing}

\begin{document}

\title{A Rapid Microlensing Event in the Q0957+561 A,B Gravitational Lens
System}

\author{Wesley N. Colley \altaffilmark{1} and
        Rudolph E. Schild \altaffilmark{2}}

\altaffiltext{1}{Dept. of Astronomy, University of Virginia, P.O. Box 3818,
     Charlottesville, VA 22903, e-mail: wnc5c@virginia.edu} 

\altaffiltext{2}{Harvard-Smithsonian Center for Astrophysics, 60 Garden Street,
Cambridge MA 02138}

\begin{abstract}

We re-analyze brightness data sampled intensively over 5 nights at two
epochs separated by the quasar lens time delay, to examine the nature of
the observed microlensing. We find strong evidence for a microlensing event
with an amplitude of 1\% and a time scale of twelve hours. The existence of
such rapid microlensing, albeit at low amplitude, imposes constraints on
the nature of the quasar and of the baryonic dark matter.

\end{abstract}


\section{Introduction}

The claimed existence of rapid microlensing in the Q0957+561 A,B
gravitational lens system has continued to challenge
astrophysics. Observationally, the low amplitude of the effect, measured to
be approximately only 0.01 magnitudes on a timescale of days (Schild
1999, Table 3; Schild \& Thomson 1993) requires careful photometry for
detection and study of the effect. Theoretically, such rapid microlensing
is not accommodated by existing models with accretion discs having
dimensions of approximately $10^{15}\mbox{cm}$ (Schmidt \& Wambsganss 1999,
Refsdal \etal\ 2000) although the new double-ring model of Schild and
Vakulik (2003) offers such rapid microlensing as a possibility provided the
baryonic dark matter is a network of planetary mass objects. Alternatives
such as orbiting dark clouds (Schechter \etal\ 2003; Wyithe \& Loeb 2003)
or bright points (Gould \& Miralde-Escude 1997) in the accretion disc
have not been advanced with detailed models or simulations because they
would produce strong periodic effects not observed and because they do not
produce the equal positive and negative events found in the data (Schild
1999).

Because such rapid microlensing, albeit at low amplitude, imposes such
severe constraints on models of the quasar and the baryonic matter
distribution in the halo of the lens galaxy G1, a determined effort to
confirm it with a sustained surveillance of the system for a 10-day
campaign by observatories circling the globe has been undertaken (Colley
\etal\ 2002, 2003). Unfortunately the system was relatively quiescent
during the campaign period, and no convincing microlensing event was
detected, although a significantly improved time delay value of 417.07 days
was measured. The campaign also demonstrated convincingly that precise
brightness measurements could be made for the system with available
photometric techniques. This conclusion has also been confirmed by other
authors (Colley \& Schild, 1999; Ovaldsen 2002; Ovaldsen \etal\ 2003).
Other photometries have been undertaken, but have lacked the precision or
intensive time sampling to detect the microlensing (Gil-Merino \etal\ 2001;
Schmidt and Wambsganss, 1998)

The purpose of the present report is to present evidence for a significant
microlensing effect in published data (Colley \& Schild 1999). The data
were previously analyzed for time delay, but because data for a single
observatory can only cover a fraction of a day and produce systematic
effects known to bias the delay calculation, full analysis of the data
required an independent time delay determination; it was this dataset that
spawned the 'round-the-clock monitoring program mentioned previously. With
the improved time delay value now available, we return to the earlier data
set and find evidence for a microlensing event having a time scale of half
a day and an amplitude of 0.01 magnitudes (one percent).

\section{Return to a High Precision Data Set}

In an earlier report, Colley \& Schild (1999) showed reductions of
all-night Q0957 brightness monitoring with the 1.2m Mt Hopkins reflector,
for datasets taken at two epochs (December 1994 and February 1996)
separated by the quasar time delay of 417 days.  Possible systematic errors
in the photometric procedure have been exhaustively discussed in Colley \&
Schild (1998, 1999) who found that the statistical errors of photometric
detection dominated over the extremely small systematic errors. This has
been confirmed in our monitoring program and reduction of CCD photometry
from 12 observatories (Colley \etal\ 2002, 2003). In the discussion of the
data from Colley \& Schild (1999) to follow, we adopt error estimates from
the original report.

Colley \& Schild (1999) measured a time-delay of $417.4$ days from the 1994
and 1996 datasets alone; in their Fig.~7 they applied that delay to the
image B record and overlayed it on top of the image A record to produce a
combined brightness record that exhibited nearly continuous fluctuations
with an amplitude of order 30 mmag over the 6 day observation window.
However, on night $\mbox{JD}-2449705.8$, an apparently significant
discrepancy was noted, which would imply microlensing on a time scale of
approximately a day. This did not spawn a detailed statistical treatment as
important a discovery as it might be, because the known systematic effects
in the time delay determination could too easily be responsible in some
unknowable way. It did, however, spawn the 10-night QuOC-Around-The Clock
monitoring campaign which did not find comparable events but did produce an
unbiased value of the time delay and also produced confidence in the
ability to convert astronomical CCD images into photometry with accuracy
sufficient to define the photometric behavior to the required precision.

\begin{figure}
\epsscale{0.9}
\plotone{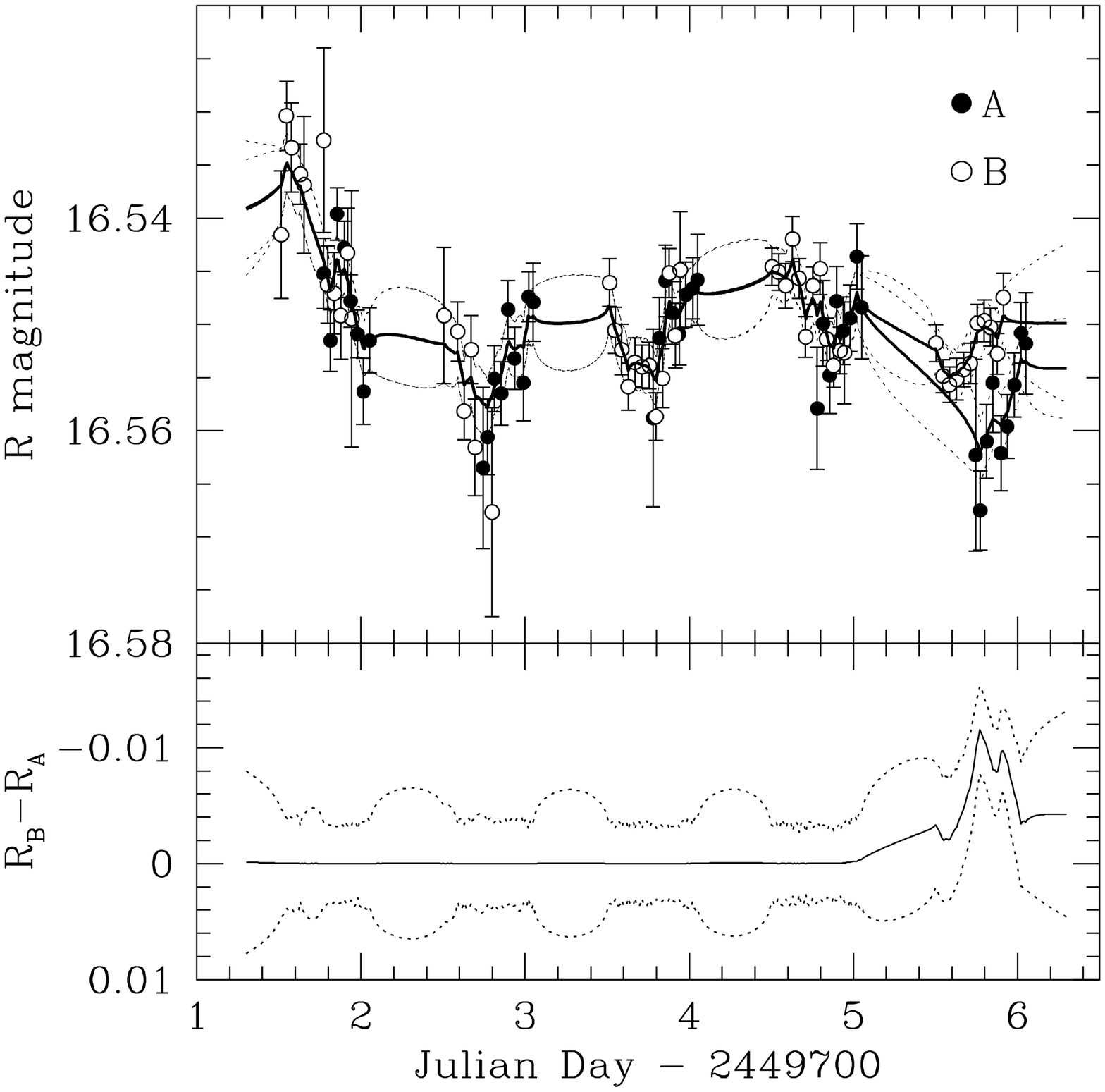}
\caption{R magnitude brightness (upper) for the Q0957 gravitational lens
images, with 1994.9 A image (filled dots) and 1996.1 data (open circles)
for a 417.07 day lag and with Julian dates given for image A. Dotted curves
show the 1-$\sigma$ confidence limits from the previously determined
structure function at the interpolation intervals. For JD 2449706, all data
points for image B are found to significantly differ from image A. (lower)
The $\mbox{B}-\mbox{A}$ brightness difference, inferred to indicate rapid
microlensing, is shown as the solid line, with confidence contours again
calculated from the measured bars and from the calculated confidence limits
for the measured quasar variability structure function.}
\label{f1}
\end{figure}

We show in Fig.~\ref{f1} a new plot of the same data for our improved
417.07 time delay value.  For such a nearly integral time delay, the data
now overlap sufficiently to allow a firm microlensing conclusion. We
consider that the data for $\mbox{JD}-2449700 = 5.8$ show a compelling
disagreement between the first arriving A image brightness and the second
arriving B brightness.  Note that for no time delay value near 417 days
would the data be in agreement, and only for time delays near 417.5 days
would there be a problem because of course the data then do not actually
overlap, but even in this case a probable microlensing would be implied.

In re-plotting the Fig.~\ref{f1} data for the new time delay, we have again
used the PRH method (Press \etal\ 1991) but with a small caveat. Because
the data set indicate the significant microlensing event as already noted,
we have excluded the data for night 2449706 in determining the optimum
brightness offset between the two images. In other words, for observing
dates fixed by the measured time delay, we use the data for the first four
nights only, to determine the magnitude difference that minimizes the
$\chi^{2}$ difference of the two brightness records.  A much more detailed
statistical treatment follows in the next section.

\section{The profile of an observed microlensing event}

In Fig.~\ref{f1} (bottom) we adopt standard precepts and determine the
microlensing profile by subtracting from the 1994.9 A image brightness the
1996.1 B image brightness, using the PRH method to define the acceptable
1-sigma allowed brightness interval from the observed brightness and the
measured structure function. In this way we are able to determine the outer
limits to allowed brightness to be compared with the measured brightness of
the other image.  Thus in Fig.~\ref{f1} (bottom) we show the time delay
corrected brightness difference which is interpreted as a microlensing
signal.  Whether the event is a brightening of image B or dimming of image
A is debatable.  Perhaps it is slightly more conservative to guess that
image A has become fainter, which would allow image B to remain fairly
constant between nights 4.5 and 5.5 without invoking a cancelling QSO
fluctuation and microlensing event.

In this illustration we see that the measured brightness differences is
significantly greater than that permitted by the errorbars and a
no-microlensing hypothesis---all data points exceed the permitted error
limits significantly.  Moreover, the brightness departures are in the
expected form of an ``event'' where the microlensing causes first a nearly
monotonic decrease in the image A brightness (or increase in the image B
brightness), and then an increase (or decrease in B) back into agreement
with the other image.  The amplitude and duration of the event were
approximately one percent in half a day.

\begin{figure}
\epsscale{0.9}
\plotone{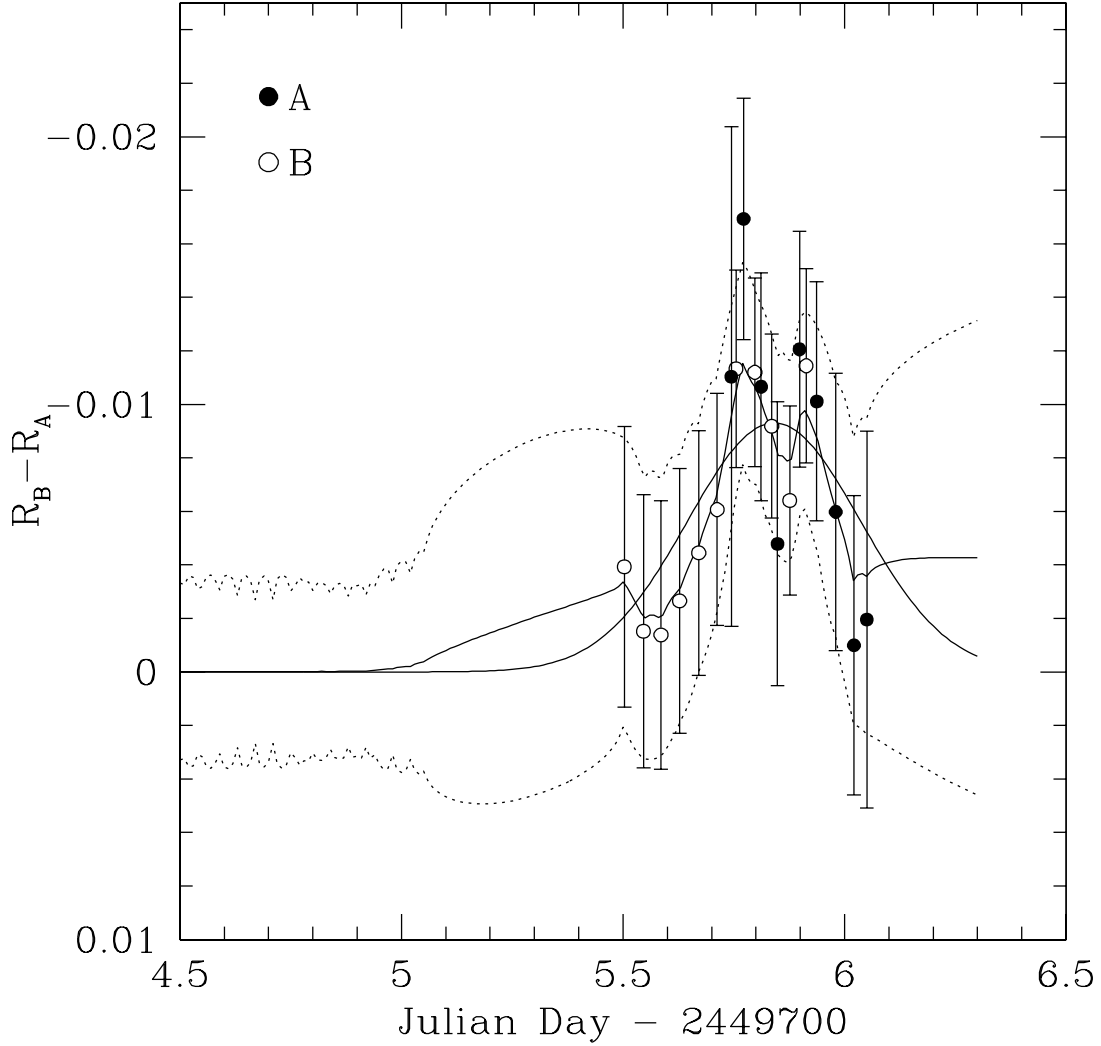}
\caption{Close-up of the microlensing event in Fig.~\ref{f1}.  
A smooth analytical profile 
shown as a heavy solid line is fitted to the PRH ``snake'', also shown as a
heavy solid line with 1-$\sigma$ outer limits shown as fine
lines. Individual data points with error bars determined as described in
section 3 are shown for image A(1994.9, filled dots) and image B(1996.1,
open circles).  The significance of the event, relative to a hypothesis
that no microlensing was observed, is within a few parts per billion of
100\%. ($\chi^2 = 80.4$ for 20 d.o.f.).}
\label{f2}
\end{figure}

We show in Fig.~\ref{f2} a close-up of the event with a fiducial Gaussian
curve fitted to the event.  The Gaussian fit has an amplitude of 0.93\%
(0.0093 mag) and a FWHM of 0.46 days.  We have overplotted points which
reflect the differences between the image A and B $R$-magnitudes, using the
PRH snake as an interpolation method, so the filled symbols are
$R_{B,int}-R_A$, and the open symbols are $R_B-R_{A,int}$.  Each errorbar
reflects the intrinsic photometric error of the un-interpolated quantity,
added in quadrature to the error reported from the PRH method for the
interpolated quantity.  There are 20 points in all, (11 B points and 9 A
points), and they have a combined $\chi^2$ relative to the zero-line of
80.4 (for 20 d.o.f), inconsistent with the zero-line at 3 parts in a
billion.  Looking only at the 12 points where there is direct overlap of
data ($5.7 < \mbox{JD} - 2449700 < 5.96$), the $\chi^2$ value (for 12
d.o.f.) relative to the zero-line is 77.0, inconsistent with the zero-line
at 1 part in a billion.  So, most of the significance lies in points which
directly overlap.  For the simple Gaussian curve fit to the points, those
$\chi^{2}$ values decrease to 9.2 and 7.4 for all points, and overlapping
points, respectively.

We may further test the significance of the event by testing whether or not
removing the last night was really more sensible than, for instance,
removing any of the other nights.  We carry this out by computing the
PRH-$\chi^2$ (with optimum magnitude offset for image B), leaving each of
the 5 nights in turn.  The PRH-$\chi^2$ for the whole data set is 94.8 for
94 degrees of freedom (d.o.f).  Leaving out nights \{1, 2, 3, 4, 5\} yields
values of \{72.0, 80.3, 76.8, 78.3, 64.5\}, with d.o.f. \{75, 79, 74, 74,
74\}.  One way to look at this by considering the change in the $\chi^2$ of
the fit relative to the change in d.o.f., $\Delta \chi^2 / \Delta N_{dof}$.
That statistic for each night is $\Delta \chi^2 / \Delta N_{dof}$ = \{1.20,
0.965, 0.903, 0.826, 1.52\}.  The last night's value is the highest of all,
showing greater significance for removing the last night than removing any
of the other nights.  The $F$-test significances of leaving out each night,
relative to leaving out none are \{58.5\%, 48.2\%, 44.6\%, 41.1\%,
74.3\%\}.  Removing any of the first four nights yields very little
improvement or actually some degradation in the quality of the fit, but
removing the last night significantly improves the fit.

\section{Conclusion and Discussion}

Previously published data have been combined with our new Q0957 time delay
value (Colley \etal\ 2003) to uncover an important rapid microlensing
event. Previously this topic has been mired in a logical conundrum; because
rapid microlensing is observed, it is extremely difficult to determine an
accurate time delay, but an accurate time delay is needed to obtain
convincing microlensing evidence. This conundrum has now been broken with
the Colley \etal\ (2003) intensive monitoring campaign, which has
sufficiently low microlensing to produce a credible time delay with sub-day
precision.

We have found convincing evidence for a low amplitude (1 percent) rapid
event with half-day duration. We provide an analytical approximation for
this event, with the view that this profile might be useful for other
analyses of the rapid microlensing. Note that Schild (1999) had previously
shown from wavelet analysis that brightness fluctuations in both images of
the Q0957 system have characteristic amplitudes of a percent on time scales
of a day or two. That analysis did not prove that such events were
predominantly microlensing; the present data suggest that the Q0957 system
shows both microlensing and intrinsic quasar brightness fluctuations of 1\%
amplitude on day time scales.

The existence of both the rapid quasar fluctuations and the rapid
microlensing fluctuations are challenging to astrophysics. For such
intrinsic quasar brightness fluctuations to exist, the quasar should have
structure on scales of half a light day ($1 \times 10^{15}\mbox{cm}$),
whereas accretion disc sizes estimated for this quasar have been (6 --
10)$\times 10^{15}\mbox{cm}$ (Refsdal \etal\ 2000), and Schmidt and
Wambsganss (1999) have ruled out sizes smaller than $10^{15}\mbox{cm}$ if
the baryonic dark matter has planetary mass.  The diameter of the innermost
stable orbit for an accretion disc around a $3 \times 10^9 M_{\odot}$ black
hole has been given by Colley \etal\ (2003) as $10^{16}\mbox{cm}$, or
approximately 3 light days. Converting all the diameter measurements to
light days, and noting that the observed times should all be increased by a
($1+z$) factor of 2.4 to allow for the cosmological redshift, we conclude
that quasar structure modeled as a simple accretion disc cannot easily
accommodate the rapid brightness fluctuations observed.

If rapid microlensing fluctuations such as this one are common, as
suggested by Schild \& Thomson (1993), they place strong constraints on
lensing models, because not only must the quasar structure be smaller than
implied in the models above, but also, the microlensing cusp pattern must
be sufficiently fine to produce such rapid events.  Attempts to explain
rapid fluctuations with (only) stellar mass deflectors in the lens galaxy
include QSO models with orbital bright blobs (Gould and Miralde-Escude,
1997) or dark clouds (Schechter \etal\ 2002, Wyithe \& Loeb 2002), but fail
because strong periodicity, and asymmetrical profiles with preponderantly
positive (brightening) or negative profiles would result.  However a
double-ring model of quasar structure advanced by Schild and Vakulik (2003)
has been demonstrated which can produce rapid microlensing fluctuations
without orbiting structures. The simulations available with microlensing by
a $0.1 M_{\odot}$ star and 90\% baryonic dark matter objects of planetary
mass produces microlensing events of 10-day duration. Reducing the dark
matter population in the simulation from $10^{-5} M_{\odot}$ to $10^{-7}
M_{\odot}$ and sharpening the inner ring structure of the model would
probably produce events of the duration observed.

We conclude that the detection of a rapid, low amplitude microlensing event
in the Q0957 dataset implies strong constraints on the nature of the
quasar's structure, and on the nature of the baryonic dark matter. Because
of the high optical depth of the Q0957 quasar to microlensing, exceeding
one, study of low amplitude microlensing fluctuations promises to impose
severe constraints on quasar structure and the nature of the baryonic dark
matter.

\newpage

\end{document}